\documentclass[%
  apl,
  amsmath,amssymb,
  reprint,%
  floatfix,
]{revtex4-1}

\usepackage{graphicx}%
\usepackage{dcolumn}%
\usepackage{bm}%

\usepackage[utf8]{inputenc}
\usepackage[T1]{fontenc}
\usepackage{mathptmx}
\usepackage{etoolbox}

\usepackage{braket}
\usepackage{siunitx}
\DeclareSIUnit\electronVolt{eV}
\usepackage{tikz}
\usepackage{multirow}
\usetikzlibrary{quantikz2}

\newcommand{\lowtemp}{\SI{350}{\milli\kelvin}}
\newcommand{\midtemp}{\SI{500}{\milli\kelvin}}
\newcommand{\hightemp}{\SI{750}{\milli\kelvin}}
\newcommand{\tTwostar}{\ensuremath{T_2^*}}
\newcommand{\tTwoecho}{\ensuremath{T_2^{\text{echo}}}}

\newcommand{\tOne}{\ensuremath{T_1}}

\makeatother
\begin{document}

\preprint{APL/123-QED}

\title[
Entangled States at 750mK]{
Persistence of Entangled States and High Fidelity Quantum Gate Operations \\ in Si/SiGe Spin Qubits at High Temperature}

\author{S. Amitonov}
\affiliation{Equal1 Laboratories}
\author{A. Apr\`{a}}
\affiliation{Equal1 Laboratories}
\author{M. Asker}
\affiliation{Equal1 Laboratories}
\author{R. Bals}
\affiliation{Equal1 Laboratories}
\author{B. Barry}
\affiliation{Equal1 Laboratories}
\author{I. Bashir}
\affiliation{Equal1 Laboratories}
\author{E. Blokhina}
\affiliation{Equal1 Laboratories}
\author{P. Giounanlis}
\affiliation{Equal1 Laboratories}
\author{M.~Harkin}
\affiliation{Equal1 Laboratories}
\author{P. Hanos-Puskai}
\affiliation{Equal1 Laboratories}
\author{I. Kriekouki}
\affiliation{Equal1 Laboratories}
\author{D. Leipold}
\affiliation{Equal1 Laboratories}
\author{M. Moras}
\affiliation{Equal1 Laboratories}
\author{N. Murphy}
\affiliation{Equal1 Laboratories}
\author{N. Petropoulos}
\affiliation{Equal1 Laboratories}
\author{C. Power}
\affiliation{Equal1 Laboratories}
\author{A.~Sammak}
\affiliation{Equal1 Laboratories}
\author{N. Samkharadze}
\affiliation{Equal1 Laboratories}
\author{A. Semenov}
\affiliation{Equal1 Laboratories}
\author{A. Sokolov}
\affiliation{Equal1 Laboratories}
\author{D. Redmond}
\affiliation{Equal1 Laboratories}
\author{C. Rohrbacher}
\affiliation{Equal1 Laboratories}
\author{X. Wu}
\affiliation{Equal1 Laboratories}

\begin{abstract}
  We characterize single- and two-qubit operations in a SiGe quantum dot array, from the perspective of its quantum information processing
  capabilities. The analysis includes rigorous randomized benchmarking of single- and
  two-qubit gates, SPAM characterization, and Bell's state tomography, which are all basic
  functionality required for universal quantum computation.  To assess compatibility with integrated cryogenic electronics, we evaluate qubit performance at \lowtemp, \midtemp, and
  \hightemp, with high fidelity single and two qubit operations. The highest temperature, \hightemp, falls within the realistic thermal budget for practical integrated cryogenic electronics and represents the highest operating temperature reported for this qubit platform.
\end{abstract}

\maketitle

Spin qubits in gate-defined quantum dots have emerged as a leading platform for
scalable quantum computing, offering long coherence times, small footprints,
and compatibility with industrial semiconductor processes \cite{Philips2022,
noiri2022, Mills2022, Huang2024, Zwerver2022, weinstein2023}.  Among these,
Si/SiGe qubits with micromagnets stand out, with SiGe spacers isolating qubits
from noisy oxide interfaces and micromagnets enabling fast driving, addressable
readout, and efficient CZ gates.  SiGe spin qubit systems achieve qubit
operation and readout fidelities beyond the error correction
threshold~\cite{xue2022, Mills2022, noiri2022}. They support the largest
universal multi-spin qubit processors and demonstrate significant advances in
operation and quantum information processing~\cite{Philips2022, Takeda2024,
noiri2022}.  Additionally, promising methods for intermediate and long-distance
qubit connectivity through single electron shuttlers~\cite{seidler2022,
DeSmet2024} and microwave resonators~\cite{dijkema2023}, underline their
potential for modular and scalable architectures.

Scaling spin qubits to utility scale requires integration of control
electronics~\cite{staszewski2022, bashir2024, pauka2021, xue2021}, ideally
co-packaged with qubits in a single System-on-Chip to minimize latency and
leverage high-density 3D interconnects. However, this integration introduces
significant thermal constraints~\cite{bashir2024}, imposing a temperature limit
below which operation becomes impractical. Therefore, it is crucial to raise the
operating temperature of these devices while ensuring that individual qubit
error rates are as low as possible.

In this work, we systematically investigate and benchmark the performance of two physical qubits in a quantum dot array in Si/SiGe at three temperatures, \lowtemp, \midtemp, and \hightemp.
Our goal is to determine the explicit temperature dependence of the coherence time $\tTwostar{}$, quantum gate fidelities, and entangled-state fidelities, thereby identifying the upper temperature limit for reliable operation of this qubit platform. We demonstrate qubit operation up to \hightemp, which, to our knowledge, represents the highest operating temperature reported for Si/SiGe spin qubits~\cite{undseth2023} {\color{black} and the only results on 1 and 2 qubit randomized benchmarking together with Bell's state tomography at this temperature for Si/SiGe}.
To this end, we calibrate and benchmark a universal native gate set
for a two-qubit processor, and perform Bell state tomography.
The extracted single- and two-qubit fidelities show only a modest degradation up to \midtemp, followed by a more pronounced decline at \hightemp. At higher temperatures, state preparation and measurement (SPAM) errors increase due to reduced sensitivity of the single-electron transistor used for readout. After correcting for SPAM, we obtain Bell states with fidelities exceeding {\color{black}85\%} at all temperatures.
Demonstrating high-fidelity Bell-state generation, tomography, and SPAM correction at elevated temperatures marks an important step toward the practical realization of scalable quantum information processing in the SiGe platform.

Devices are fabricated on an undoped $^{28}$Si/SiGe heterostructure
featuring a \SI{7}{\nano\meter} strained $^{28}$Si quantum well, grown on a strain-relaxed
Si$_{0.7}$Ge$_{0.3}$ buffer layer. The quantum well is separated from
the surface by a \SI{30}{\nano\meter} thick Si$_{0.7}$Ge$_{0.3}$ spacer, terminated
by a Si-rich passivation layer~\cite{Esposti2022APL}. The gate stack
consists of 3 layers of Ti:Pd metallic gates
isolated from each other by Al$_2$O$_3$ dielectric layers, deposited
using atomic layer deposition. A~ferromagnetic Ti:Co layer
on top of the gate stack creates a local magnetic field gradient for
qubit addressing and manipulation.
{
  \color{black}
  The external magnetic field was set to \SI{29}{\milli\tesla}.
  The heterostructure was specifically tuned for higher valley splitting~\cite{Esposti2022APL}; for such devices, valley splitting was found to be $\gtrsim \SI{0.1}{\milli\electronVolt}$~\cite{Esposti2024}, which corresponds to a temperature range $T \gtrsim \SI{1}{\kelvin}$.
}
Further details of device
fabrication methods can be found in~\cite{lawrie2020quantum}.

\begin{figure*}[ht]
  \centering
  \includegraphics[width=\linewidth]{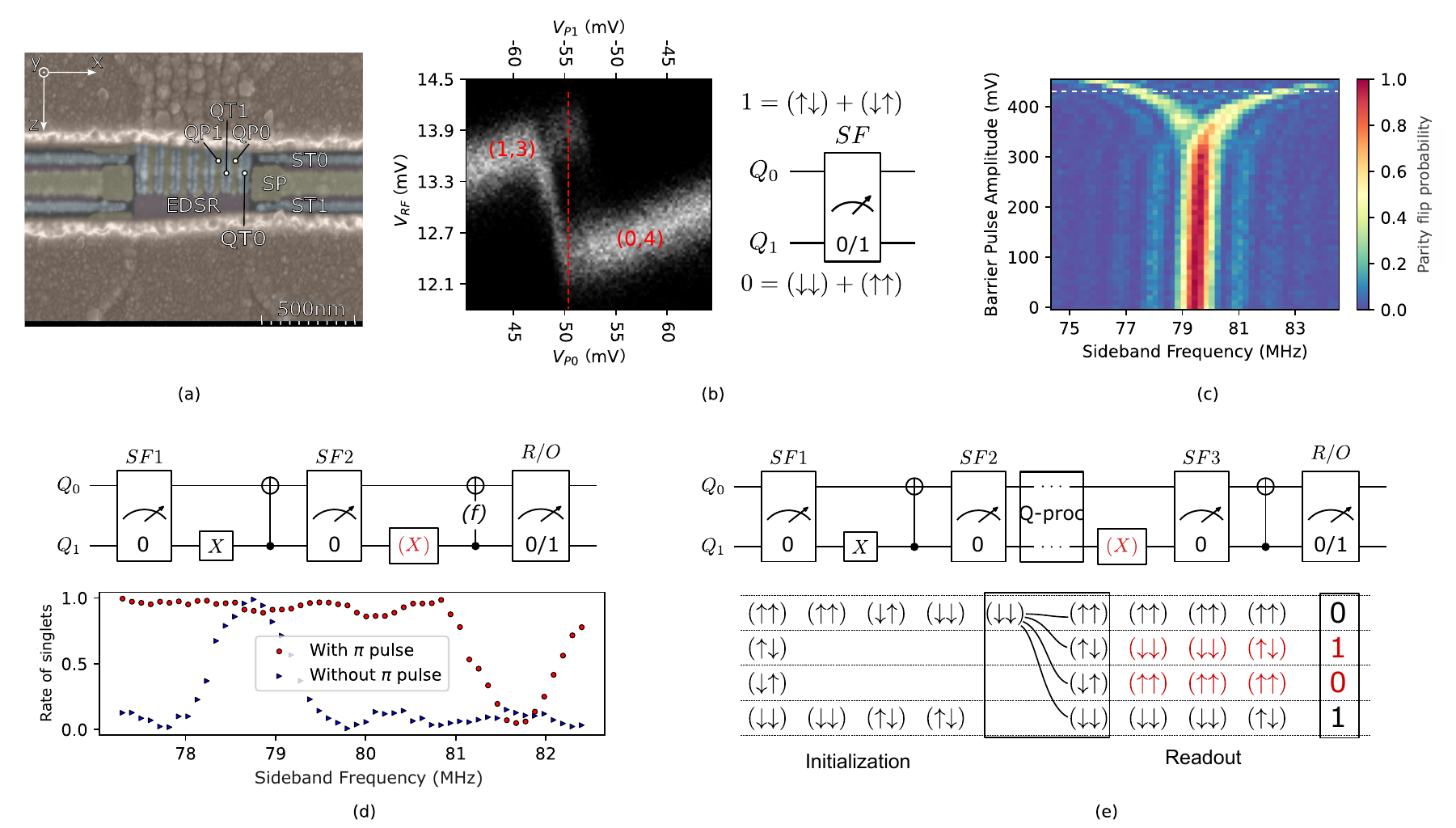}
  \caption{\textbf{(a)}  Scanning electron micrograph of a 6-quantum dot device
    nominally identical to the one measured.
    \textbf{(b)} On the left: histogram of the charge sensor signal as a
    function of the detuning $\epsilon$, scanning perpendicularly to
    interdot charge
    transition. The {\color{black}red} dashed line identifies the readout point,
    where parity readout is enabled.
    Parity readout is used for a state filter (SF; on the right), which
    discards antiparallel spin states.
    \textbf{(c)} Even to odd parity flip probability.
    The pulse on the interdot barrier activates the exchange coupling, exhibiting the typical
    spin funnel shape (for more details on this experiment see~\cite{Petta2005}).
    \textbf{(d)} Demonstration of CROT gate. Following a
    double initialization stage, as explained in main text, a CROT is applied on
    target qubit $Q_0$ with control qubit $Q_1$. We show the
    percentage of singlets
    measured with and without a $\pi$ rotation on the control qubit $Q_1$. The
    virtual barrier amplitude is approximately {\color{black}\SI{420}{\milli\volt}}, as indicated
    by dashed line
    in (c).
    \textbf{(e)} Initialization and readout of two qubit state. We
    apply an additional
    state filtering (SF3) followed by a CROT operation for the
    two-qubit readout.
    This allows to distinguish between two states with same parity,
    as shown in the
    schematics. By applying an additional X rotation on control $Q_1$ before the
    readout stage, SF3, we can map odd states into even states, thus
    enabling the
    readout of odd states.
    {\color{black}All state filterings here refer to the same quantum process.}
  }
  \label{fig:Fig1}
\end{figure*}

A false-colored SEM image of our device is presented in Fig.\,\ref{fig:Fig1}\,(a).
{\color{black}
  For details on the device fabrication, please refer to the Supplementary materials.
}
The structure comprises a six-quantum-dot array with two single-electron transistors (SETs) located at the edges for charge-state detection. We use two qubits in this study. The array is formed by plunger (QP) gates  to control the dots chemical potentials and tunnelling QT gates to control inter-dot and dot-sensor tunnel couplings.
Quantum
dots are formed at the energy minima, energy levels of which are controlled
by modifying the voltage applied on the plunger gates, while the height and
width of the barriers between adjacent dots are controlled by adjusting the
biases of the tunneling gates. The details of the experimental setup used for the measurement as well as the methodoly used to tune the quantum dot are given in the Supplementary material.

Quantum dot transitions in the array are probed with an RF-SET
\cite{Reflecto_review}, where the sensor dot is controlled by plunger SP, and
tunneling with reservoirs by tunneling gates  ST0, ST1
(see Fig.\,\ref{fig:Fig1}\,(a)).
A high impedance superconducting inductor is connected to
the accumulation gate of the RF-SET \cite{PhysRevApplied.16.014057}.
We accumulate a double quantum dot (DQD) underneath the plunger gates QP0 and
QP1. In the following we refer to the corresponding qubits as $Q_0$ and
$Q_1$.  After optimizing the charge sensor dot, the tunnel barriers controlled
by gates QT0, QT1  are then tuned to bring the DQD into a regime suitable
for Pauli Spin Blockade (PSB).
{\color{black}
  The qubit driving is facilitated by electric dipole spin resonance mechanism induced by the transversal field of the micromagnet. The quantum dots are confined from the bottom side by EDSR line (see Fig.~\ref{fig:Fig1}(a)) used for driving and screening, and from the top by a dedicated screening electrode (not seen on the image) shaped in a similar way to the EDSR.
}

Fig.\,\ref{fig:Fig1}\,(b) shows the histogram of the charge sensor signal
as a function of interdot detuning $\epsilon = V_{\text{QP}0} - V_{\text{QP}1}$.
The red dashed line indicates the  chosen readout point. At this readout
point we can observe parity-dependent tunneling. This allows measurement of the
parity of the spin states of $Q_0$ and $Q_1$. In the
circuit diagrams we label even states as $0$ and odd states as $1$.

Spin manipulation is enabled by the gradient of a micromagnet  field that
generates a "synthetic" spin-orbit coupling~\cite{Slanting_field}. The
micromagnet field is maximized on qubit $Q_0$ and becomes lower as we move
further along the qubit array. Fig.\,\ref{fig:Fig1}\,(c) shows the effect of exchange coupling on the qubit resonance frequency.
{\color{black}
  For a double dot, in the presence of the magnetic field detuning along the quantization axis the states $\ket{\uparrow, \downarrow}$ and $\ket{\downarrow, \uparrow}$ have different energies. The line before the bifurcation in Fig.~1(c) shows the resonance frequency of $\ket{\uparrow, \downarrow} \to \ket{\downarrow, \downarrow}$ and $\ket{\downarrow, \uparrow} \to \ket{\uparrow, \uparrow}$ transitions (or $\ket{\uparrow, \downarrow} \to \ket{\uparrow, \uparrow}$ and $\ket{\downarrow, \uparrow} \to \ket{\downarrow, \downarrow}$; depending upon the ``target'' qubit of choice), which do not depend upon the state of the other qubit.
  By controlling the interdot tunnelling barrier (by means of controlling gate voltages) we can move the electrons closer to each other, and as a consequence increase the exchange interaction. This moves the resonance frequencies of the above transitions asymmetrically by the magnitude of $J$ coupling, resulting in two resonance lines. Now the resonance frequencies of the transitions depend upon the state of the other qubit, which allows us to perform
  conditional operations on the two qubit system.
  Because of the non-linear change in magnetic field detuning along the quantization axis and the barrier pulse amplitude vs interelectron distance dependence, we see a non-linear behavior of the resonance lines after the bifurcation.
}

The controlled-rotation(CROT) operation is shown in Fig. 1(d). We first initialize the system in the $\ket{\downarrow,\downarrow}$ state using a two-stage state-filtering sequence~\cite{Huang2024}. This involves performing a Pauli spin blockade (PSB) operation and applying a state filter (SF1) to discard the antiparallel spin states. A $\pi$ rotation is then applied to qubit Q1, followed by a CROT operation on Q0 and a second state filter (SF2). Next, we activate the exchange interaction and sweep the frequency of the microwave excitation. By preparing the control qubit Q1 in either the $\ket{\downarrow}$ or $\ket{\uparrow}$ state, we selectively activate one of the two branches of the spin-exchange funnel. The resonance curves of the two branches approach 100$\%$, demonstrating the effectiveness of the initialization of the two-qubit state.

Figure 1(e) illustrates the extension of the two-qubit initialization scheme to include full two-qubit state readout using PSB. The sequence begins with the same two-stage initialization procedure. After quantum processing, the first PSB step results in partial information loss, preventing discrimination between the $\ket{\downarrow,\downarrow}$ and $\ket{\downarrow, \uparrow}$ states due to fast relaxation from $\ket{\downarrow, \uparrow}$ to $\ket{\uparrow, \downarrow}$ at the readout point \cite{Philips2022}. To address this limitation, each experiment is performed twice: once using the standard sequence, and once with an additional X gate applied to Q1 after the intended quantum processing. This approach allows reconstruction of all possible final two-qubit states after quantum processing by consistently selecting the parallel spin states in the final state filter (SF3). Looking ahead, a real time feedback technique \cite{Philips2022, Kobayashi2023} will allow to initialize quasi-deterministically all 4 two-spin states, thereby enhancing the operational speed of the quantum processor.

\begin{figure}[t]
  \centering
  \begin{tikzpicture}[font=\sffamily]
    \node[anchor=south west,inner sep=0] (img) at (0,0)
    {\includegraphics{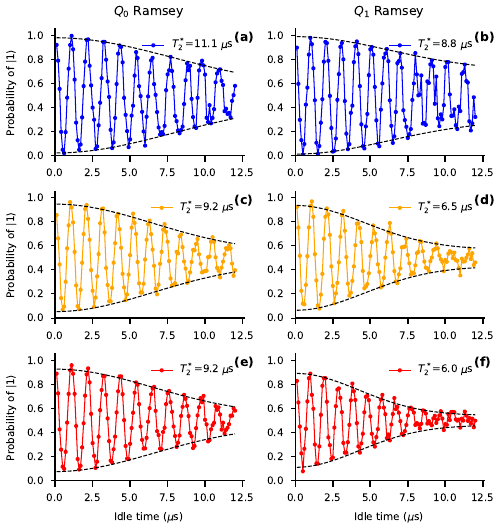}};
    \node[scale=0.65] at (3.5,6.55) {T\,=\,350~mK};
    \node[scale=0.65] at (7.5,6.55) {T\,=\,350~mK};
    \node[scale=0.65] at (3.5,3.8)  {T\,=\,500~mK};
    \node[scale=0.65] at (7.5,3.8)  {T\,=\,500~mK};
    \node[scale=0.65] at (3.5,1.05) {T\,=\,750~mK};
    \node[scale=0.65] at (7.5,1.05) {T\,=\,750~mK};
  \end{tikzpicture}
  \caption{Free evolution coherence \tTwostar{} times measurements with the Ramsey protocol at three different temperatures. We add a waiting time-dependent phase shift to the last $\pi/2$ pulse with 1~MHz artificial detuning to highlight the decay envelope.
    \textbf{(a)}, \textbf{(b)} Ramsey interference fringes for $Q_0$ and $Q_1$ at $\lowtemp$ (blue);
    \textbf{(c)}, \textbf{(d)} at $\midtemp$ (orange);
    \textbf{(e)}, \textbf{(f)} at $\hightemp$ (red).
    \tTwostar{} dephasing times are extracted by fitting the envelopes of the Ramsey experiment data, indicated by black dashed lines.
  }
  \label{fig:Fig2_ramsey}
\end{figure}

\begin{figure}[ht]
  \centering
  \includegraphics{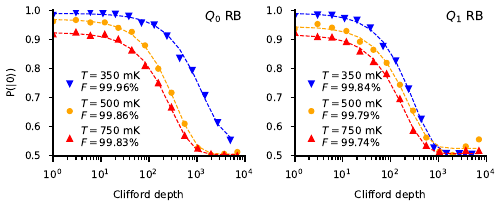}
  \caption{
    Clifford randomized benchmarking for both qubits
    at temperatures of \lowtemp{} (blue, $\blacktriangledown$), \midtemp{} (orange, $\bullet$), and \hightemp{} (red, $\blacktriangle$).
    Each datapoint represents an average return probability after
    applying 15 random Clifford sequences per Clifford depth.
    Fidelity data is given in the figures and summarised in Table~\ref{tab:performance_summary}.
  }
  \label{fig:benchmarking_q1}
\end{figure}

Next, we characterize qubit performance at three different temperatures: $T=\lowtemp$, $\midtemp$ and
$\hightemp$.
Fig.~\ref{fig:Fig2_ramsey} shows Ramsey fringes used to extract coherence times of the two qubits. A change in \tTwostar{} is observed with increasing temperature in Q0 and Q1, dropping by approximately 17\% and 32\%, respectively. This indicates an  increase in charge noise between \lowtemp{} and \hightemp{} as it is believed to be the dominant contributor to
decoherence in spin-orbit coupled system~\cite{micromagnet_noise_coupling}. The coherence time can be extended with the Hahn echo protocol (Table~\ref{tab:performance_summary}).

\begin{figure}[t]
  \centering
  \begin{tikzpicture}[font=\sffamily]
    \node[anchor=south west,inner sep=0] (img) at (0,0)
    {\includegraphics{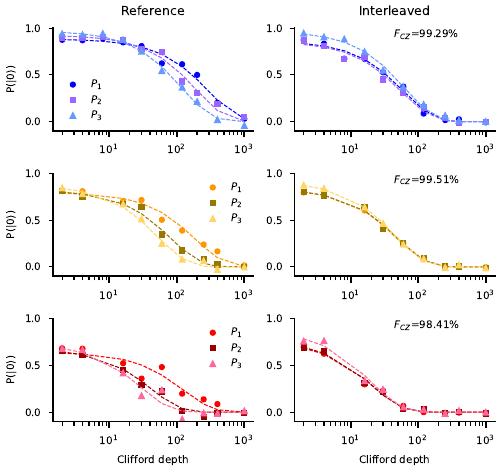}};
    \node[scale=0.65] at (3.5, 7.3)  {T\,=\,350~mK};
    \node[scale=0.65] at (7.25, 7.1)   {T\,=\,350~mK};
    \node[scale=0.65] at (1.65,3.5)  {T\,=\,500~mK};
    \node[scale=0.65] at (7.25,4.6)  {T\,=\,500~mK};
    \node[scale=0.65] at (1.65,1.05) {T\,=\,750~mK};
    \node[scale=0.65] at (7.25,2.15) {T\,=\,750~mK};
  \end{tikzpicture}
  \caption{
    Benchmarking $CZ$ gate fidelities using ICRB at three different
    temperatures: \lowtemp{} (blue); \midtemp{} (orange); \hightemp{} (red).
    The left and right columns show results for, respectively, the reference and
    interleaved benchmarking sequences for $P_1$, $P_2$, $P_3$.
    Each datapoint represents an average return probability after
    applying 10 random Clifford sequences per Clifford depth.
    The extracted fidelities are summarised in Table~\ref{tab:performance_summary}.
  }
  \label{fig:Fig2}
\end{figure}

\begin{figure*}[t]
  \centering
  \begin{tikzpicture}[font=\sffamily]
    \node[anchor=south west,inner sep=0] (img) at (0,0)
    {\includegraphics{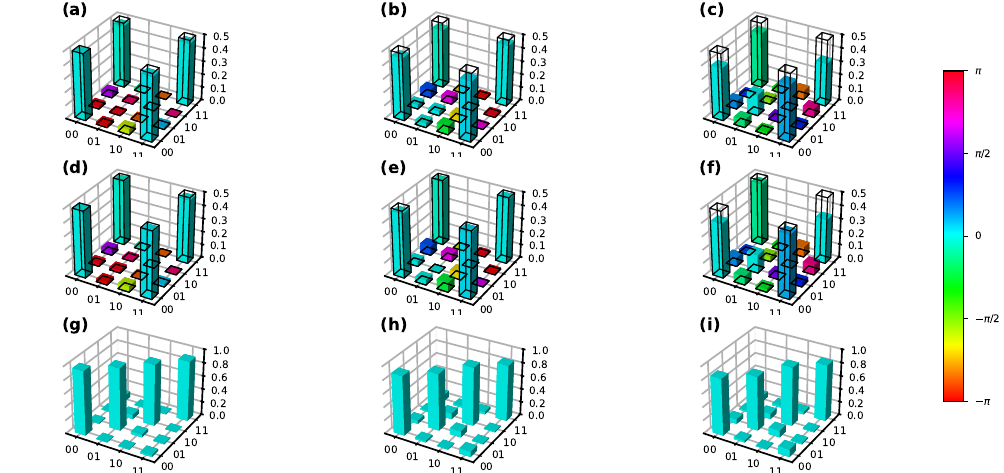}};
    \node[scale=0.75] at (3.0, 8) {T\,=\,350~mK};
    \node[scale=0.75] at (8.4,8)  {T\,=\,500~mK};
    \node[scale=0.75] at (13.8,8) {T\,=\,750~mK};
  \end{tikzpicture}
  \caption{(a)--(c) Reconstructed via tomography density matrices for the Bell state
    $\frac{1}{\sqrt{2}}(\ket{00} + \ket{11})$,
    (d)--(f) corresponding SPAM-corrected density matrices, and
    (g)--(i) measured SPAM matrices.
    The vertical axis represents the real component, color indicates phase.
    The black wireframe for the Bell state results indicates where the
    real component would be ideally.
    The columns from left to right represent the results for the temperatures
    $T=\lowtemp$ ((a),(d),(g)), \midtemp{} ((b), (e), (h)), and \hightemp{}
    ((c), (f), (i)), respectively.
    Fidelity and concurrence values before and after SPAM correction are
    given in Table~\ref{tab:performance_summary}.
  }
  \label{fig:tomography_bell}
\end{figure*}

To extract the fidelities of one qubit gates we use the standard Clifford randomized benchmarking
(RB)~\cite{Magesan2012}. The protocol was performed using the
$\{I, X(\pm \frac{\pi}{2}),Z(\pm \frac{\pi}{2}), X(\pm \pi), Z(\pm \pi)\}$
native gate-set, with the virtual $Z$ gate implemented as a
phase shift~\cite{mckay_efficient_2017} to compile 1-qubit Clifford gates.
The virtual $Z$ gate allows
for a near-perfect implementation of the $Z$ gate with zero duration,
greatly improving the one-qubit gate fidelity.
Each Clifford gate was
compiled into a circuit with an average of 0.83 one-qubit $X$ rotations.
We sample $N=15$ random Clifford gate sequences for each depth $d\leq 10^3$ and take
$M=10^4$ shots per measurement.
The  values of $P(\ket{0})$
post-selected and averaged over the $N$ Clifford gate sequences
for each depth
are fitted with $W = Ab^d + C$, where $A$, $b$ and $C$ are fitting
parameters, and the fidelity is calculated as $F = (1+b)/2$.
The experimental results and their fits are shown in
Fig.~\ref{fig:benchmarking_q1}.

Our native entangler is a controlled-Z ($CZ$) operation that we implement by pulsing on the barrier gate to turn on the exchange and performing single qubit phase rotation~\cite{watson2018programmable} (see Supplementary Material). We characterize the $CZ$ fidelity using the
Interleaved Character Randomized Benchmarking procedure (ICRB)~\cite{Xue2019}.
The standard interleaved randomized benchmarking
technique~\cite{Magesan2012_inter}, when extended to two qubit gates,
results in relatively deep
($3.9$ one qubit non-virtual gates and $1.48$ $CZ$ gates per Clifford),
circuits in our native gate set,  $\{I, X(\pm \frac{\pi}{2}),Z(\pm
\frac{\pi}{2}), X(\pm \pi), Z(\pm \pi), CZ\}$.
This leads to a fast decay in the resulting benchmarking curve giving
low predictions of two-qubit gate fidelity with large uncertainty.
We avoid this by using ICRB which allows one to calculate the fidelity
of a two qubit gate using finite groups other than the Clifford group
(for the technical details of the method see~\cite{Helsen2019}).
With ICRB one chooses a benchmarking group and a character group.
Whilst the benchmarking group plays the same role as in the standard
RB, it is the additional character group that allows freedom of
choice in the benchmarking group (for the standard RB, the
benchmarking group is fixed to $n$-qubit Clifford group $\mathrm{Clif}_n$).
By careful choice of the benchmarking and character groups within the
ICRB one can avoid noise accumulation by exploiting high-fidelity
gates and concentrating on the target gate.
For spin qubits it is convenient to choose
a product of two one-qubit Clifford groups $\mathrm{Clif}_1 \otimes
\mathrm{Clif}_1$, and a two-qubit Pauli group $\mathcal{P}_2$ as a
character group ~\cite{Xue2019}.
The experimental results and their processing are shown in Fig.~\ref{fig:Fig2}.

It is important to note that even without full optimization for each temperature the 1 and 2 qubit fidelity results are {\color{black}comparable~\footnote{{\color{black} Though in the paper~\cite{xue2022} the authors have used the Gate-set tomography technique to recover the information on the noise channels, to calculate gate fidelity they have used an equation for an average fidelity obtained after twirling the channel into the depolarizing channel~\cite{emerson2005}, which is equivalent to fidelity obtained using the randomized benchmarking procedure.}}} to the state of the art for Si/SiGe~\cite{Camenzind2025paper, xue2022}.  %

To verify entanglement in our system, we perform full state
tomography on the Bell state
$\ket{\Phi^+} = \frac{1}{\sqrt{2}}(\ket{00} + \ket{11}) $.
Our reconstruction of the density matrix is based on its the Pauli
decomposition
$
\rho = \sum_{\sigma \in P_2} c_\sigma \sigma,
$
where $P_2 = \{ I, \sigma_x, \sigma_y, \sigma_z \}^{\otimes 2}$ and
$c_\sigma = \langle \sigma \rangle/4$ are a set of all two-qubit
Pauli matrix combinations and a normalized expectation value of a
two-qubit Pauli operator $\sigma$, respectively. The latter we find
by measuring expectation values of all 16 two-qubit Pauli strings.
The resulting density matrices $T=\lowtemp$, $T=\midtemp$ and
$\hightemp$ are visualized in
Fig.~\ref{fig:tomography_bell}.

\begin{table*}[ht]
  \centering
  \begin{tabular}{|l|c|c|c|c|c|c|}
    \hline
    Temp.         & \multicolumn{2}{c|}{\lowtemp}                                                     & \multicolumn{2}{c|}{\midtemp}                                             & \multicolumn{2}{c|}{\hightemp} \\ \hline
    Qubit         &  $Q_0$                                      & $Q_1$                               & $Q_0$                               & $Q_1$                               & $Q_0$                               & $Q_1$  \\ \hline
    Freq.         &  {\SI{5.047}{\giga\hertz}}          & {\SI{4.991}{\giga\hertz}}   &    {\SI{5.047}{\giga\hertz}}     &     {\SI{4.991}{\giga\hertz}}   & {\SI{5.047}{\giga\hertz}}   & {\SI{4.991}{\giga\hertz}}    \\ \hline
    $T_1$         & $\gtrsim$ \SI{134}{\milli\second}           & $\gtrsim$ \SI{134}{\milli\second}   &                $\gtrsim$    \SI{73}{\milli\second}                 &         $\gtrsim$   \SI{73}{\milli\second}                        & $\gtrsim$ \SI{41}{\milli\second}    & $\gtrsim$ \SI{41}{\milli\second}     \\ \hline
    $\tTwostar$   & {11.1} \SI{}{\micro\second}          & {8.8} \SI{}{\micro\second} & {9.2} \SI{}{\micro\second} & {6.5} \SI{}{\micro\second} & {9.2} \SI{}{\micro\second} & {5.9} \SI{}{\micro\second}  \\ \hline
    $\tTwoecho$   & {43.3 \SI{}{\micro\second} }         & {43.6 \SI{}{\micro\second} }   &   {36.4 \SI{}{\micro\second} }    & {38.3 \SI{}{\micro\second} }  & {31.9 \SI{}{\micro\second} } & {32 \SI{}{\micro\second} }    \\ \hline
    X90           & {\SI{140}{\nano\second}}           & {\SI{140}{\nano\second}}   &     {\SI{140}{\nano\second}}                                 &   {\SI{140}{\nano\second}}                                   & {\SI{140}{\nano\second}}    & {\SI{140}{\nano\second}}     \\ \hline
    CZ180         & \multicolumn{2}{c|}{\SI{200}{\nano\second}}   &    \multicolumn{2}{c|}{\SI{200}{\nano\second}}         & \multicolumn{2}{c|}{\SI{200}{\nano\second}}     \\ \hline
    1Q $F_{Clif}$ & {99.96\%}                           & {99.84\%}                   &                       {99.86\%}              &       {99.79\%}               & {99.83\%}                   & {99.74\%} \\ \hline
    2Q $F_{CZ}$   & \multicolumn{2}{c|}{{99.30\%}}                                             & \multicolumn{2}{c|}{{99.51\%}}                                                    & \multicolumn{2}{c|}{{98.41\%}}                                                  \\ \hline
    Bell fid.     & \multicolumn{2}{c|}{0.9590}                                                       & \multicolumn{2}{c|}{0.9028}                                               & \multicolumn{2}{c|}{0.7576}                                             \\ \hline
    Bell fid.(s)  & \multicolumn{2}{c|}{0.9794}                                                       & \multicolumn{2}{c|}{0.9675}                                               & \multicolumn{2}{c|}{0.8192}                                             \\ \hline
    Bell Con      & \multicolumn{2}{c|}{0.9124}                                                       & \multicolumn{2}{c|}{0.7564}                                               & \multicolumn{2}{c|}{0.4404}                                             \\ \hline
    Bell Con.(s)  & \multicolumn{2}{c|}{0.9442}                                                       & \multicolumn{2}{c|}{0.8229}                                               & \multicolumn{2}{c|}{0.4869}                                             \\ \hline
  \end{tabular}
  \caption{\label{tab:performance_summary} Summary of qubit metrics.
    Bell fid. (s) refers to the fidelity of the Bell state without
    (and with SPAM correction).
  }
\end{table*}

Table~\ref{tab:performance_summary} summarizes the measured qubit properties and two-qubit gate performance across all temperatures. To isolate the effect of temperature, qubit frequencies and gate durations were kept constant. As expected, the relaxation rate increases with temperature, yet $\tOne$ remains long at 41 ms even at \hightemp{}, making relaxation negligible on experimental timescales. Both the free-evolution ($\tTwostar$) and echo ($\tTwoecho$) coherence times decrease with temperature, consistent with enhanced charge noise coupling through spin–orbit interaction~\cite{micromagnet_noise_coupling}. Similarly, one-qubit Clifford and two-qubit CZ gate fidelities degrade at higher temperatures.
A minor increase in CZ fidelity at \midtemp{} ($F_{CZ}=99.51\%$)
compared to \lowtemp{} ($F_{CZ}=99.30\%$) is attributed to recalibration uncertainty {\color{black} (see Supplementary Material for more information)}. SPAM-corrected Bell- and CZ-state fidelities remain high at \midtemp{} but drop sharply at \hightemp{}, suggesting an optimal operating temperature around \midtemp{} for our current technology. Further extension to higher temperatures can be achieved by reducing charge noise through optimized fabrication within a production-grade \SI{300}{\milli\meter} process.

Temperature impact is most noticeable in readout visibility,
primarily due to the limitations of traditional SETs, which rely on
the sharpness of the Fermi level in the reservoirs. To address these
limitations at higher temperatures, alternative charge sensors, such
as gate sensors with superconducting inductors~\cite{zheng2019} and
double dot SETs~\cite{Huang2021}, offer promising solutions by sensing interdot transitions directly. In the future, we will focus on extending the operational temperature
range of charge sensors in qubit arrays and integrating the qubits
with our fifth-generation control chip. The control chip supports a
programmable qubit control system
and RF synthesis engine, ARM based sequencing
and scheduling core, all designed to work at compatible cryogenic
temperature with integrated qubits or external SiGe qubits dies in a
3D package configuration.

\begin{acknowledgments}
  We acknowledge the Netherlands Organisation for Applied Scientific
  Research (TNO) and its Quantum Information Technology Test (QITT)   group for their valuable support.
\end{acknowledgments}

\section*{Data Availability Statement}
The data and code used to create the figures in this study can be found at \url{http://dx.doi.org/10.5281/zenodo.18763785}.

\bibliography{references}

@article{watson2018programmable,
  title={A programmable two-qubit quantum processor in silicon},
  author={Watson, TF and Philips, SGJ and Kawakami, Erika and Ward, DR and Scarlino, Pasquale and Veldhorst, Menno and Savage, DE and Lagally, MG and Friesen, Mark and Coppersmith, SN and others},
  journal={nature},
  volume={555},
  number={7698},
  pages={633--637},
  year={2018},
  publisher={Nature Publishing Group UK London}
}

@article{Camenzind2025paper,
  title = {Simultaneous {{High-Fidelity Single-Qubit Gates}} in a {{Spin Qubit Array}}},
  author = {Wu, Yi-Hsien and Camenzind, Leon C. and B{\"u}tler, Patrick and Jin, Ik Kyeong and Noiri, Akito and Takeda, Kenta and Nakajima, Takashi and Kobayashi, Takashi and Scappucci, Giordano and Goan, Hsi-Sheng and Tarucha, Seigo},
  year = 2025,
  number = {arXiv:2507.11918},
  eprint = {2507.11918},
  primaryclass = {quant-ph},
  publisher = {arXiv},
  doi = {10.48550/arXiv.2507.11918}
}

@article{emerson2005,
  title = {Scalable Noise Estimation with Random Unitary Operators},
  author = {Emerson, Joseph and Alicki, Robert and {\.Z}yczkowski, Karol},
  year = 2005,
  journal = {Journal of Optics B: Quantum and Semiclassical Optics},
  volume = {7},
  number = {10},
  pages = {S347},
  issn = {1464-4266},
  doi = {10.1088/1464-4266/7/10/021}
}

@article{weinstein2023,
  title={Universal logic with encoded spin qubits in silicon},
  author={Weinstein, Aaron J. and Reed, Matthew D. and Jones, Aaron M. and Andrews, Reed W. and Barnes, David and Blumoff, Jacob Z. and Euliss, Larken E. and Eng, Kevin and Fong, Bryan H. and Ha, Sieu D. and Hulbert, Daniel R. and Jackson, Clayton A. C. and Jura, Michael and Keating, Tyler E. and Kerckhoff, Joseph and Kiselev, Andrey A. and Matten, Justine and Sabbir, Golam and Smith, Aaron and Wright, Jeffrey and Rakher, Matthew T. and Ladd, Thaddeus D. and Borselli, Matthew G.},
  journal={Nature},
  volume={615},
  pages={817--822},
  year={2023},
  publisher={Nature Publishing Group},
  doi={10.1038/s41586-023-05782-0},
  url={https://doi.org/10.1038/s41586-023-05782-0},
  note={19k accesses, 40 citations, 117 Altmetric}
}

@article{mckay_efficient_2017,
  title     = {Efficient {Z} gates for quantum computing},
  volume    = {96},
  copyright = {https://link.aps.org/licenses/aps-default-license},
  issn      = {2469-9926, 2469-9934},
  url       = {https://link.aps.org/doi/10.1103/PhysRevA.96.022330},
  doi       = {10.1103/PhysRevA.96.022330},
  language  = {en},
  number    = {2},
  urldate   = {2024-11-28},
  journal   = {Phys. Rev. A},
  author    = {McKay, David C. and Wood, Christopher J. and Sheldon, Sarah and Chow, Jerry M. and Gambetta, Jay M.},
  month     = aug,
  year      = {2017},
  pages     = {022330},
}

@article{Magesan2012_inter,
  title     = {Efficient Measurement of Quantum Gate Error by Interleaved Randomized Benchmarking},
  author    = {Magesan, Easwar and Gambetta, Jay M. and Johnson, B. R. and Ryan, Colm A. and Chow, Jerry M. and Merkel, Seth T. and da Silva, Marcus P. and Keefe, George A. and Rothwell, Mary B. and Ohki, Thomas A. and Ketchen, Mark B. and Steffen, M.},
  journal   = {Phys. Rev. Lett.},
  volume    = {109},
  issue     = {8},
  pages     = {080505},
  numpages  = {5},
  year      = {2012},
  month     = {Aug},
  publisher = {American Physical Society},
  doi       = {10.1103/PhysRevLett.109.080505},
  url       = {https://link.aps.org/doi/10.1103/PhysRevLett.109.080505},
}

@article{Helsen2019,
  title   = {A new class of efficient randomized benchmarking protocols},
  author  = {Helsen, J. and Xue, X. and Vandersypen, L.M.K. and Wehner, S.},
  journal = {npj Quantum Inf.},
  volume  = {5},
  pages   = {71},
  year    = {2019},
  doi     = {10.1038/s41534-019-0182-7},
  url     = {https://doi.org/10.1038/s41534-019-0182-7},
}

@article{staszewski2022,
  title     = {Cryogenic Controller for Electrostatically Controlled Quantum Dots in 22-nm Quantum SoC},
  author    = {Staszewski, Robert Bogdan and Esmailiyan, Ali and Wang, Hongying and Koskin, Eugene and Giounanlis, Panagiotis and Wu, Xutong and Koziol, Anna and Sokolov, Andrii and Bashir, Imran and Asker, Mike and others},
  journal   = {IEEE Open Journal of the Solid-State Circuits Society},
  volume    = {2},
  pages     = {103--121},
  year      = {2022},
  publisher = {IEEE},
}

@article{bashir2024,
  title     = {Monolithically Integrated Quantum Dots in a 22-nm Fully Depleted Silicon-on-Insulator Process Operating at 3 K},
  author    = {Bashir, Imran and Sokolov, Andrii and Wu, Xutong and Giounanlis, Panagiotis and Petropoulos, Nikolaos and Leipold, Dirk and Asker, Mike and Esmailiyan, Ali and Andrade-Miceli, Dennis and Haenlein, Hans-Christoph and others},
  journal   = {International Journal of Circuit Theory and Applications},
  year      = {2024},
  publisher = {Wiley Online Library},
}

@article{xue2021,
  title     = {CMOS-based cryogenic control of silicon quantum circuits},
  author    = {Xue, Xiao and Patra, Bishnu and van Dijk, Jeroen PG and Samkharadze, Nodar and Subramanian, Sushil and Corna, Andrea and Paquelet Wuetz, Brian and Jeon, Charles and Sheikh, Farhana and Juarez-Hernandez, Esdras and others},
  journal   = {Nature},
  volume    = {593},
  number    = {7858},
  pages     = {205--210},
  year      = {2021},
  publisher = {Nature Publishing Group UK London},
}

@article{Esposti2024,
  title     = {Low disorder and high valley splitting in silicon},
  author    = {Esposti, Davide Degli and Stehouwer, Lucas E. A. and G\"{u}l, \"{O}nder and Samkharadze, Nodar and D\'{e}prez, Corentin and Meyer, Marcel and Meijer, Ilja N. and Tryputen, Larysa and Karwal, Saurabh and Botifoll, Marc and Arbiol, Jordi and Amitonov, Sergey V. and Vandersypen, Lieven M. K. and Sammak, Amir and Veldhorst, Menno and Scappucci, Giordano},
  journal   = {npj Quantum Information},
  volume    = {10},
  pages     = {32},
  year      = {2024},
}

@article{Zwerver2022,
  author  = {A. M. J. Zwerver and T. Krähenmann and T. F. Watson and L. Lampert and H. C. George and R. Pillarisetty and S. A. Bojarski and P. Amin and S. V. Amitonov and J. M. Boter and R. Caudillo and D. Correas-Serrano and J. P. Dehollain and G. Droulers and E. M. Henry and R. Kotlyar and M. Lodari and F. Lüthi and D. J. Michalak and B. K. Mueller and S. Neyens and J. Roberts and N. Samkharadze and G. Zheng and O. K. Zietz and G. Scappucci and M. Veldhorst and L. M. K. Vandersypen and J. S. Clarke},
  title   = {Qubits made by advanced semiconductor manufacturing},
  journal = {Nature Electronics},
  volume  = {5},
  pages   = {184--190},
  year    = {2022},
  month   = {March},
  doi     = {10.1038/s41928-022-00727-9},
  url     = {https://doi.org/10.1038/s41928-022-00727-9},
}

@article{DeSmet2024,
  author  = {Maxim De Smet and Yuta Matsumoto and Anne-Marije J. Zwerver and Larysa Tryputen and Sander L. de Snoo and Sergey V. Amitonov and Amir Sammak and Nodar Samkharadze and Önder Gül and Rick N. M. Wasserman and Maximilian Rimbach-Russ and Giordano Scappucci and Lieven M. K. Vandersypen},
  title   = {High-fidelity single-spin shuttling in silicon},
  journal = {arXiv preprint arXiv:2406.07267},
  year    = {2024},
  month   = {June},
  url     = {https://doi.org/10.48550/arXiv.2406.07267},
  note    = {15 pages, 15 figures},
}

@article{zheng2019,
  title={Rapid gate-based spin read-out in silicon using an on-chip resonator},
  author={Zheng, Guoji and Samkharadze, Nodar and Noordam, Marc L. and Kalhor, Nima and Brousse, Delphine and Sammak, Amir and Scappucci, Giordano and Vandersypen, Lieven M. K.},
  journal={Nature Nanotechnology},
  volume={14},
  pages={742--746},
  year={2019},
  publisher={Nature Publishing Group},
  doi={10.1038/s41565-019-0483-1},
  url={https://doi.org/10.1038/s41565-019-0483-1},
  note={Published: 08 July 2019, Author Correction: 25 November 2019}
}

@article{dijkema2023,
  title={Two-qubit logic between distant spins in silicon},
  author={Dijkema, Jurgen and Xue, Xiao and Harvey-Collard, Patrick and Rimbach-Russ, Maximilian and de Snoo, Sander L. and Zheng, Guoji and Sammak, Amir and Scappucci, Giordano and Vandersypen, Lieven M. K.},
  journal={arXiv preprint arXiv:2310.16805},
  year={2023},
  url={https://doi.org/10.48550/arXiv.2310.16805},
  note={17 pages, 9 figures}
}

@article{Huang2021,
  title = {A High-Sensitivity Charge Sensor for Silicon Qubits above 1 K},
  author = {Huang, Jonathan Yue and Lim, Wee Han and Leon, Ross C. C. and Yang, Chih Hwan and Hudson, Fay E. and Escott, Christopher C. and Saraiva, Andre and Dzurak, Andrew S. and Laucht, Arne},
  journal = {ACS Nano},
  year = {2021},
  month = {May},
  volume = {15},
  issue = {5},
  pages = {8097--8104},
  doi = {10.1021/acsnano.1c02463},
  publisher = {American Chemical Society},
  copyright = {2021 American Chemical Society}
}

@article{pauka2021,
  title     = {A cryogenic CMOS chip for generating control signals for multiple qubits},
  author    = {Pauka, SJ and Das, K and Kalra, R and Moini, A and Yang, Y and Trainer, M and Bousquet, A and Cantaloube, C and Dick, N and Gardner, GC and others},
  journal   = {Nature Electronics},
  volume    = {4},
  number    = {1},
  pages     = {64--70},
  year      = {2021},
  publisher = {Nature Publishing Group},
}

@article{undseth2023,
  title     = {Hotter is easier: unexpected temperature dependence of spin qubit frequencies},
  author    = {Undseth, Brennan and Pietx-Casas, Oriol and Raymenants, Eline and Mehmandoost, Mohammad and M{\k{a}}dzik, Mateusz T and Philips, Stephan GJ and De Snoo, Sander L and Michalak, David J and Amitonov, Sergey V and Tryputen, Larysa and others},
  journal   = {Physical Review X},
  volume    = {13},
  number    = {4},
  pages     = {041015},
  year      = {2023},
  publisher = {APS},
}

@article{xue2022,
  title     = {Quantum logic with spin qubits crossing the surface code threshold},
  author    = {Xue, Xiao and Russ, Maximilian and Samkharadze, Nodar and Undseth, Brennan and Sammak, Amir and Scappucci, Giordano and Vandersypen, Lieven MK},
  journal   = {Nature},
  volume    = {601},
  number    = {7893},
  pages     = {343--347},
  year      = {2022},
  publisher = {Nature Publishing Group UK London},
}

@article{seidler2022,
  title={Conveyor-mode single-electron shuttling in Si/SiGe for a scalable quantum computing architecture},
  author={Seidler, Inga and Struck, Tom and Xue, Ran and Focke, Niels and Trellenkamp, Stefan and Bluhm, Hendrik and Schreiber, Lars R.},
  journal={npj Quantum Information},
  volume={8},
  pages={100},
  year={2022},
  publisher={Nature Publishing Group},
  doi={10.1038/s41534-022-00620-y},
  url={https://doi.org/10.1038/s41534-022-00620-y},
  note={Published: 30 August 2022}
}

@article{Mills2022,
  author  = {Adam R. Mills  and Charles R. Guinn  and Michael J. Gullans  and Anthony J. Sigillito  and Mayer M. Feldman  and Erik Nielsen  and Jason R. Petta },
  title   = {Two-qubit silicon quantum processor with operation fidelity exceeding 99\%},
  journal = {Science Advances},
  volume  = {8},
  number  = {14},
  pages   = {eabn5130},
  year    = {2022},
  doi     = {10.1126/sciadv.abn5130},
  url     = {https://www.science.org/doi/abs/10.1126/sciadv.abn5130},
  eprint  = {https://www.science.org/doi/pdf/10.1126/sciadv.abn5130},
}

@article{Magesan2012,
  title     = {Characterizing quantum gates via randomized benchmarking},
  author    = {Magesan, Easwar and Gambetta, Jay M. and Emerson, Joseph},
  journal   = {Phys. Rev. A},
  volume    = {85},
  issue     = {4},
  pages     = {042311},
  numpages  = {16},
  year      = {2012},
  month     = {Apr},
  publisher = {American Physical Society},
  doi       = {10.1103/PhysRevA.85.042311},
  url       = {https://link.aps.org/doi/10.1103/PhysRevA.85.042311},
}

@article{noiri2022,
  title     = {Fast universal quantum gate above the fault-tolerance threshold in silicon},
  author    = {Noiri, Akito and Takeda, Kenta and Nakajima, Takashi and Kobayashi, Takashi and Sammak, Amir and Scappucci, Giordano and Tarucha, Seigo},
  journal   = {Nature},
  volume    = {601},
  number    = {7893},
  pages     = {338--342},
  year      = {2022},
  publisher = {Nature Publishing Group UK London},
}

@article{lawrie2020quantum,
  title     = {Quantum dot arrays in silicon and germanium},
  author    = {Lawrie, WIL and Eenink, HGJ and Hendrickx, NW and Boter, JM and Petit, L and Amitonov, SV and Lodari, M and Paquelet Wuetz, B and Volk, C and Philips, SGJ and others},
  journal   = {Applied Physics Letters},
  volume    = {116},
  number    = {8},
  year      = {2020},
  publisher = {AIP Publishing},
}

@article{Esposti2022APL,
  title     = {Wafer-scale low-disorder $\mathrm{2DEG}$ in $^{28}$$\mathrm{Si}/\mathrm{SiGe}$ without an epitaxial $\mathrm{Si}$ cap},
  author    = {Degli Esposti, D. and Paquelet Wuetz, B. and Fezzi, V. Lodari, M. and Sammak, A. and Scappucci, G.},
  journal   = {Applied Physics Letters},
  volume    = {120},
  number    = {18},
  year      = {2022},
  publisher = {AIP Publishing},
}

@article{Xue2019,
  title     = {Benchmarking Gate Fidelities in a $\mathrm{Si}/\mathrm{SiGe}$ Two-Qubit Device},
  author    = {Xue, X. and Watson, T. F. and Helsen, J. and Ward, D. R. and Savage, D. E. and Lagally, M. G. and Coppersmith, S. N. and Eriksson, M. A. and Wehner, S. and Vandersypen, L. M. K.},
  journal   = {Phys. Rev. X},
  volume    = {9},
  issue     = {2},
  pages     = {021011},
  numpages  = {12},
  year      = {2019},
  month     = {Apr},
  publisher = {American Physical Society},
  doi       = {10.1103/PhysRevX.9.021011},
  url       = {https://link.aps.org/doi/10.1103/PhysRevX.9.021011},
}

@article{Philips2022,
  author   = {Philips, Stephan G. J.
              and M{\k{a}}dzik, Mateusz T.
              and Amitonov, Sergey V.
              and de Snoo, Sander L.
              and Russ, Maximilian
              and Kalhor, Nima
              and Volk, Christian
              and Lawrie, William I. L.
              and Brousse, Delphine
              and Tryputen, Larysa
              and Wuetz, Brian Paquelet
              and Sammak, Amir
              and Veldhorst, Menno
              and Scappucci, Giordano
              and Vandersypen, Lieven M. K.},
  title    = {Universal control of a six-qubit quantum processor in silicon},
  journal  = {Nature},
  year     = {2022},
  month    = {Sep},
  day      = {01},
  volume   = {609},
  number   = {7929},
  pages    = {919-924},
  issn     = {1476-4687},
  doi      = {10.1038/s41586-022-05117-x},
  url      = {https://doi.org/10.1038/s41586-022-05117-x},
}

@article{Takeda2024,
  author   = {Takeda, Kenta
              and Noiri, Akito
              and Nakajima, Takashi
              and Camenzind, Leon C.
              and Kobayashi, Takashi
              and Sammak, Amir
              and Scappucci, Giordano
              and Tarucha, Seigo},
  title    = {Rapid single-shot parity spin readout in a silicon double quantum dot with fidelity exceeding 99{\%}},
  journal  = {npj Quantum Information},
  year     = {2024},
  month    = {Feb},
  day      = {13},
  volume   = {10},
  number   = {1},
  pages    = {22},
  issn     = {2056-6387},
  doi      = {10.1038/s41534-024-00813-0},
  url      = {https://doi.org/10.1038/s41534-024-00813-0},
}

@article{Huang2024,
  author   = {Huang, Jonathan Y. and Su, Rocky Y. and Lim, Wee Han and Feng, MengKe and van Straaten, Barnaby and Severin, Brandon and Gilbert, Will and Dumoulin Stuyck, Nard and Tanttu, Tuomo and Serrano, Santiago and Cifuentes, Jesus D. and Hansen, Ingvild and Seedhouse, Amanda E. and Vahapoglu, Ensar and Leon, Ross C. C. and Abrosimov, Nikolay V. and Pohl, Hans-Joachim and Thewalt, Michael L. W. and Hudson, Fay E. and Escott, Christopher C. and Ares, Natalia and Bartlett, Stephen D. and Morello, Andrea and Saraiva, Andre and Laucht, Arne and Dzurak, Andrew S. and Yang, Chih Hwan},
  title    = {High-fidelity spin qubit operation and algorithmic initialization above 1 K},
  journal  = {Nature},
  year     = {2024},
  month    = {Mar},
  day      = {01},
  volume   = {627},
  number   = {8005},
  pages    = {772-777},
  issn     = {1476-4687},
  doi      = {10.1038/s41586-024-07160-2},
  url      = {https://doi.org/10.1038/s41586-024-07160-2},
}

@article{Kobayashi2023,
  author   = {Kobayashi, T.
              and Nakajima, T.
              and Takeda, K.
              and Noiri, A.
              and Yoneda, J.
              and Tarucha, S.},
  title    = {Feedback-based active reset of a spin qubit in silicon},
  journal  = {npj Quantum Information},
  year     = {2023},
  month    = {Jun},
  day      = {01},
  volume   = {9},
  number   = {1},
  pages    = {52},
  issn     = {2056-6387},
  doi      = {10.1038/s41534-023-00719-3},
  url      = {https://doi.org/10.1038/s41534-023-00719-3},
}

@article{PhysRevApplied.16.014057,
  title = {Radio-Frequency Reflectometry in Silicon-Based Quantum Dots},
  author = {Liu, Y.-Y. and Philips, S.G.J. and Orona, L.A. and Samkharadze, N. and McJunkin, T. and MacQuarrie, E.R. and Eriksson, M.A. and Vandersypen, L.M.K. and Yacoby, A.},
  journal = {Phys. Rev. Appl.},
  volume = {16},
  issue = {1},
  pages = {014057},
  numpages = {7},
  year = {2021},
  month = {Jul},
  publisher = {American Physical Society},
  doi = {10.1103/PhysRevApplied.16.014057},
  url = {https://link.aps.org/doi/10.1103/PhysRevApplied.16.014057}
}

@article{Reflecto_review,
title = "Probing quantum devices with radio-frequency reflectometry",
author = "F. Vigneau and Federico Fedele and Anasua Chatterjee and David Reilly and F Kuemmeth and Gonzalez-Zalba, {M. F.} and Edward Laird and Natalia Ares",
year = "2023",
month = jun,
day = "30",
doi = "10.1063/5.0088229",
language = "English",
volume = "10",
journal = "Applied Physics Reviews",
issn = "1931-9401",
publisher = "American Institute of Physics Publising LLC",
number = "2",
}

@Article{Slanting_field,
author={Pioro-Ladri{\`e}re, M.
and Obata, T.
and Tokura, Y.
and Shin, Y.-S.
and Kubo, T.
and Yoshida, K.
and Taniyama, T.
and Tarucha, S.},
title={Electrically driven single-electron spin resonance in a slanting Zeeman field},
journal={Nature Physics},
year={2008},
month={Oct},
day={01},
volume={4},
number={10},
pages={776-779},
issn={1745-2481},
doi={10.1038/nphys1053},
url={https://doi.org/10.1038/nphys1053}
}

@article{Petta2005,
author = {J. R. Petta  and A. C. Johnson  and J. M. Taylor  and E. A. Laird  and A. Yacoby  and M. D. Lukin  and C. M. Marcus  and M. P. Hanson  and A. C. Gossard },
title = {Coherent Manipulation of Coupled Electron Spins in Semiconductor Quantum Dots},
journal = {Science},
volume = {309},
number = {5744},
pages = {2180-2184},
year = {2005},
doi = {10.1126/science.1116955},
URL = {https://www.science.org/doi/abs/10.1126/science.1116955},
eprint = {https://www.science.org/doi/pdf/10.1126/science.1116955}}

@article{micromagnet_noise_coupling,
    author = {Kha, Allen and Joynt, Robert and Culcer, Dimitrie},
    title = {Do micromagnets expose spin qubits to charge and Johnson noise?},
    journal = {Applied Physics Letters},
    volume = {107},
    number = {17},
    pages = {172101},
    year = {2015},
    month = {10},
    issn = {0003-6951},
    doi = {10.1063/1.4934693},
    url = {https://doi.org/10.1063/1.4934693},
    eprint = {https://pubs.aip.org/aip/apl/article-pdf/doi/10.1063/1.4934693/13149626/172101\_1\_online.pdf},
}
\newpage

\section*{Supplementary Material}

\section{Device fabrication}

Devices used in this study are fabricated by the Netherlands Organization for Applied Scientific Research (TNO) using their commercial foundry service. Devices are fabricated on an undoped $^{28}$Si/SiGe heterostructure featuring a \SI{7}{\nano\meter} strained $^{28}$Si quantum well, grown on a strain-relaxed Si$_{0.7}$Ge$_{0.3}$ buffer layer. The quantum well is separated from the surface by a \SI{30}{\nano\meter} thick Si$_{0.7}$Ge$_{0.3}$ spacer, terminated by a Si-rich passivation
layer~\cite{Esposti2022APL}. The gate stack is separated from the heterostructure by \SI{10}{\nano\meter} Al$_2$O$_3$, formed by atomic-layer deposition (ALD) at 300$^o$C. The three gate layers of the gate stack are made from Ti:Pd with thicknesses of 3:17, 3:27 and 3:\SI{37}{\nano\meter} and are patterned using electron beam lithography, electron beam evaporation and lift off. Each layer is electrically isolated from the previous layer by a \SI{5}{\nano\meter} Al$_2$O$_3$ dielectric grown by ALD. Above the three gate layers a micro magnet is fabricated from Ti:Co (5:\SI{200}{\nano\meter}). Further details of device fabrication methods can be found in~\cite{lawrie2020quantum}.

\section{Sample Packaging and Thermalization}
To ensure optimal thermal anchoring, the sample is mounted directly onto an Oxygen-Free Copper (OFC) plate. This plate is mechanically bolted to the cold finger of the dilution refrigerator. Thermal contact between the sample substrate and the OFC plate is established using silver conductive paint (SCP).

\section{Experimental setup}

\begin{figure}[h]
  \centering
  \includegraphics[width=1\columnwidth]{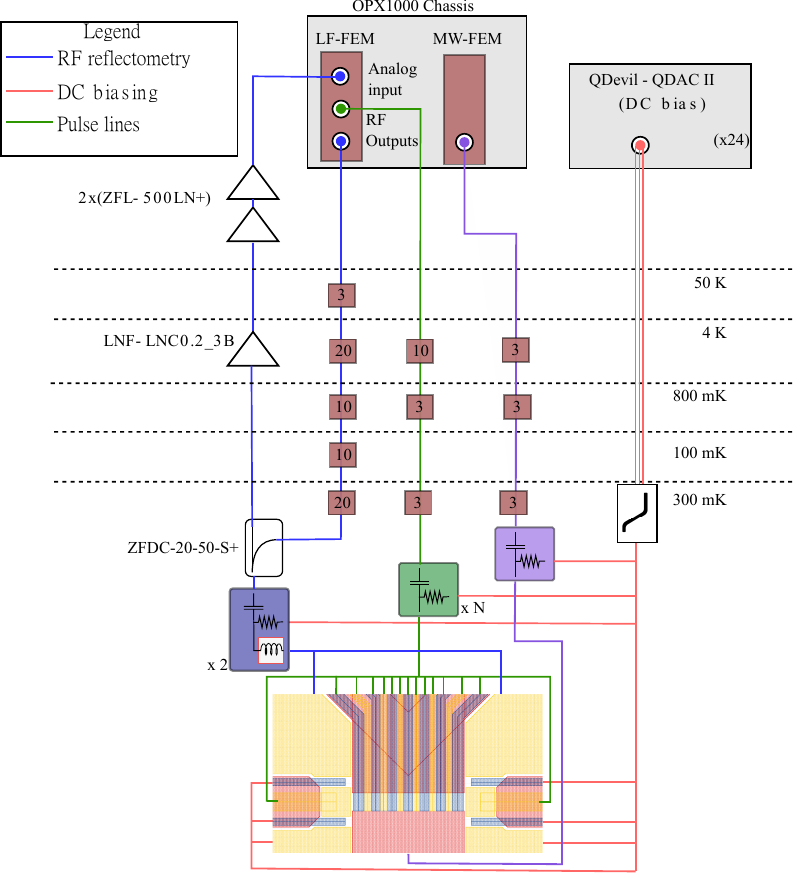}
  \caption{Schematic overview of the experimental setup inside the Bluefors dilution cryostat. The experiment is controlled using an OPX1000 chassis with one microwave module (MW-FEM) and two low-frequency module (LF-FEM).
    LF-FEMs are used for readout and baseband pulses. MW-FEM cards is used for the microwave control signal send to the EDSR line. Reflectometry tank circuit, bias-tees and sample are all located on the same PCB but are separated here for clarity. DC signals are controlled using a QDAC-II, signals are filtered by several stages of LC and RC filters on a custom made PCB thermalized at base temperature. Base temperature of the mixing chamber is 10 mK, but it is heated between 300~mK to 750~mK for the purpose of this study.
  }
  \label{fig:setup}
\end{figure}

\section{Device tuning}

\begin{figure}
  \centering
  \includegraphics[width=1\columnwidth]{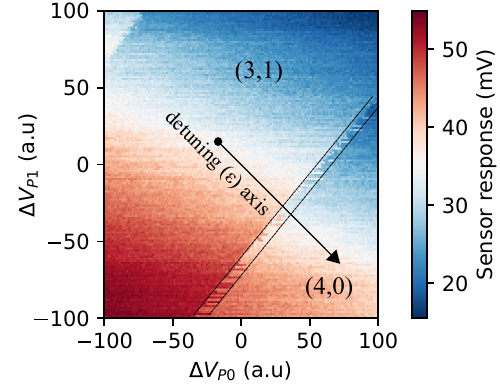}
  \caption{ Stability diagram of the (4,0)--(3,1) transition, captured using pulsed gate measurements in video mode. In our operating mode the quantum dots are decoupled from reservoirs, hence only the interdot transition is visible here. The readout window is indicated here by the dashed lines.
  }
  \label{stability}
\end{figure}

The device is operated in the (1,3) - (0,4) regime where the readout window is maximized~\cite{Philips2022}, the other dots are not loaded. To tune the device we used pulsed gate measurements in a videomode software. We first tune the dot in the correct regime using the SET as a reservoir, for ease of tuning we used virtualized gate set. Once the dot are in the correct regime, we close off the tunnel coupling to the reservoir using the QT0 gate. The resulting stability diagramm is visible in Fig.~\ref{stability} where the readout window is clearly visible.

\begin{figure}
  \centering
  \includegraphics[width=\linewidth]{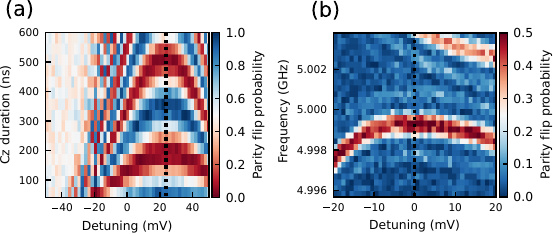}
  \caption{Scans showing the dependence of exchange with detuning $\epsilon$. In (a) we perfrom a $DCZ$ gate which consist of two $ CZ$ gate pulses interleaved with an echo-sequence at a fixed exchange pulse level set to 402 mV to fit a $CZ$ duration of \SI{200}{\nano\second}. To reduce the effect of charge noise on $CZ$ we perform the gate at the symmetry point indicated by the dashed line. In (b), we perform a frequency scan with varying detuning; here, the exchange pulse is set to \SI{420}{\milli\volt}. This is used to tune the operating point of the CROT gate used in the initialization scheme.  Again, we perform the CROT gate at the symmetry point for the selected branch indicated by the dash line.
  }
  \label{fig:2qgate}
\end{figure}

To minimize the charge noise susceptibility of the exchange interaction, the CROT and the $CZ$ gate have to be performed at their respective exchange symmetry points. To tune the CROT symmetry point we perform a frequency scan at the desired exchange level with respect to detuning as shown in Fig. \ref{fig:2qgate}(b). For $CZ$ We perform a decoupled-CZ, or $DCZ$ with respect to detuning, which consists of two exchange pulses interleaved with an echo-pulse, results are visible in Fig. \ref{fig:2qgate} (b).

\section{Additional experimental results}

\begin{figure}
  \centering
  \begin{tikzpicture}[font=\sffamily]
    \node[anchor=south west,inner sep=0] (img) at (0,0)
    {\includegraphics{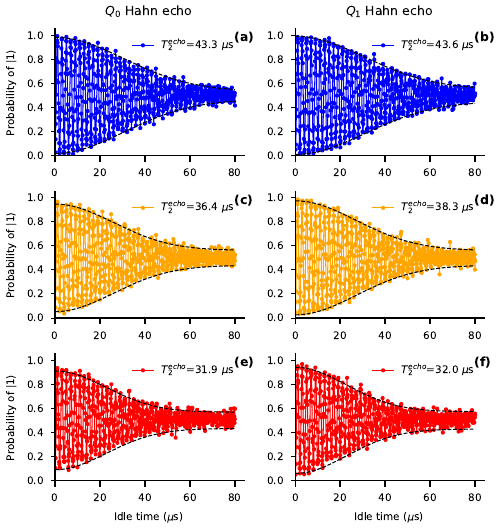}};
    \node[scale=0.65] at (3.5,6.55) {T\,=\,350~mK};
    \node[scale=0.65] at (7.5,6.55) {T\,=\,350~mK};
    \node[scale=0.65] at (3.5,3.8)  {T\,=\,500~mK};
    \node[scale=0.65] at (7.5,3.8)  {T\,=\,500~mK};
    \node[scale=0.65] at (3.5,1.05) {T\,=\,750~mK};
    \node[scale=0.65] at (7.5,1.05) {T\,=\,750~mK};
  \end{tikzpicture}
  \caption{Hahn echo results at three different temperatures:
    \lowtemp{} (blue plots); \midtemp{} (orange plots); \hightemp{} (red plots).
    The dephasing rates $\tTwoecho$ are extracted by fitting to the data,
    indicated by a black dashed line, and given in Table~\ref{tab:performance_summary}.
  }
  \label{fig:hahn_fit}
\end{figure}

\begin{figure*}
  \includegraphics{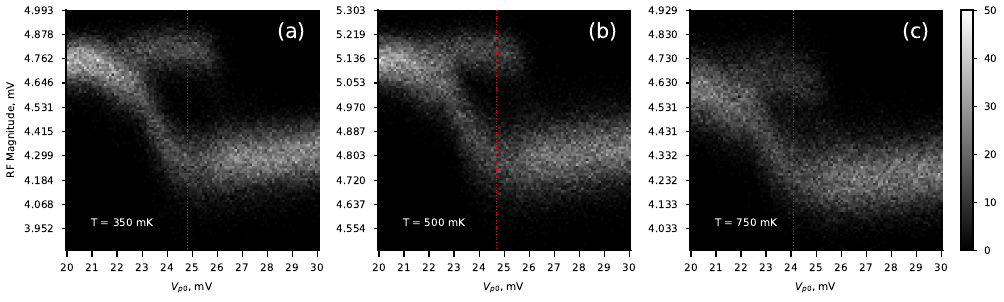}
  \label{supfig:effect_of_temp_on_readout}
  \caption{
    Readout window for each temperature \textbf{(a)} \lowtemp{}, \textbf{(b)} \midtemp{} and \textbf{(c)} \hightemp{}. Integration time is kept constant at \SI{0.2}{\milli\second}. We observe degradation of readout SNR caused by broadening of SET peak at higher temperatures.
    The red dotted line indicates the readout point used for subsequent measurements in this calibration regime.
  }
\end{figure*}

\section{Discussion on fidelity results vs recalibration uncertainty}

A large number of experimental parameters contribute to the fine-tuning of fidelity, including the resonance frequency of the LC tank circuit, the chosen integration time, the type of pulses used for qubit driving and their duration.

The optimisation of both single-qubit (1Q) and two-qubit (2Q) gate fidelities depends on multiple control parameters. In principle, two approaches can be considered. First, one could re-optimise several parameters (such as pulse duration, drive power and related settings) for each qubit at each temperature  to obtain the highest achievable fidelities. Alternatively, one may fix a subset of parameters to mimic a more realistic operating scenario in a quantum processor where, for example, the duration of a single-qubit X gate is expected to be uniform across qubits (as well as the power of the driving pulses), and then determine the achievable 1Q and 2Q fidelities under these constraints.

In this work, we opted for the latter approach and compared fidelities using a common set of fixed parameters rather than optimising each qubit individually. In principle, every qubit likely has its own optimal parameter set, and we therefore do not claim full optimisation of the measurement configuration. Instead, we claim typical repeatable fidelity under some constraints. The observed variations in fidelity should not be interpreted as systematic deviations, but instead as a consequence of applying identical, non-optimised settings across different qubits. These results also suggest that further improvements in 2Q fidelity may be possible, for example through adjustments to pulse duration or drive amplitude.

\end{document}